# Cell dynamics simulation of droplet and bridge formation within striped nano-capillaries

*Masao Iwamatsu*

Department of Physics, General Education Center,

Musashi Institute of Technology,

Setagaya-ku, Tokyo 158-8557, Japan

iwamatsu@ph.ns.musashi-tech.ac

The kinetics of droplet and bridge formation within striped nano-capillaries is studied when the wetting film grows via interface-limited growth. The phenomenological time-dependent Ginzburg-Landau (TDGL)-type model with thermal noise is used and numerically solved using the cell dynamics method. The model is two-dimensional and consists of undersaturated vapor confined within a nano-capillary made of two infinitely wide flat substrates. The surface of the substrate is chemically heterogeneous with a single stripe of lyophilic domain that exerts long-range attractive potential to the vapor molecule. The dynamics of nucleation and subsequent growth of droplet and bridge can be simulated and visualized. In particular, the evolution of the morphology from droplet or bump to bridge is clearly identified. Crucial role played by the substrate potential on the morphology of bridge of nanoscopic size is clarified. Nearly temperature-independent evolution of capillary condensation is predicted when the interface-limited growth dominates. In addition, it is shown that the dynamics of capillary condensation follows the scenario of capillary condensation proposed by Everett and Haynes three decades ago.

## I. Introduction

Vapor condenses within nano-capillary when the width of capillary is narrow and the substrate is lyophilic. Eventually condensed liquid fills whole capillary, which is known as the capillary condensation. In contrast to heterogeneous nucleation where liquid condenses from oversaturated vapor, liquid condenses from undersaturated vapor in capillary condensation due to the attractive substrate potential. Capillary condensation is so ubiquitous phenomena that it has been studied over the centuries. Although the equilibrium thermodynamics of capillary condensation is understood fairly well,[1-4] the dynamics has not received the same attention. Recently, partly due to the advances in various types of scanning microscopes,[5] information concerning the dynamics of capillary condensation has become available.[6-8] Such information is important because it is related to various geographical, biological processes as well as technological problems related to nanotechnology. Unfortunately, however, theoretical studies of the dynamics of capillary condensation are quite scarce[9-13] and are mostly simulation studies. In addition, nucleation and subsequent growth of the wetting layer has not been distinguished and analyzed separately in previous studies.[6-8]

In 1970's Everett and Haynes[14] has pointed out that the nucleation of capillary condensation is the bridge formation between the opposite walls of the capillary. Even though there have been various studies of the morphology and thermodynamic stability of liquid bridges,[10,15,16] the direct simulation so far has not confirmed the evolution of bridge[9-12] and Everett-Haynes scenario of capillary condensation. This is because the bridge is metastable.[15] Without imposing certain constraint on density fluctuation, chemical or geometrical heterogeneity, or artificial field, the direct simulation of bridge formation is difficult. Therefore, a theoretical model that can clearly separate the nucleation and growth of bridge is highly desired.

In this paper, we focus on the dynamics of nucleation of bridge formation and subsequent growth of the wetting layer to confirm the Everett-Haynes scenario[14] of bridging. We will use the standard

Ginzburg-Landau or phase-field type model.[9,10] Therefore the phase transformation occurs via the interface-limited growth of the wetting film. In order to promote the bridging, we use the chemically heterogeneous striped surface, which can act as the seed for the bridge formation. Our simulation will clearly trace the evolution of the morphology of droplet and bridge and confirm the Everett-Haynes scenario of capillary condensation.

Our article is organized as follows. In section II, the time-dependent Ginzburg-Landau (TDGL) model is introduced and the essential feature of substrate potential to the condensation is given. We also give the general expression of wall potential made by the lyophilic stripe. Numerical technique using the cell dynamics method, and the numerical results for droplet and capillary bridge formation are given in section III. Finally section IV is devoted to the conclusion.

## II. Model

**A. Time-Dependent Ginzburg-Landau (TDGL) model.** We start from the standard time-dependent Ginzburg Landau (TDGL)[9,10] evolution model:

$$\frac{\partial \psi}{\partial t} = -\frac{\delta F}{\delta \psi} \qquad (1)$$

where $\delta$ denotes the functional differentiation, $\psi$ is the non-conserved order parameter, and $F[\psi]$ is the free energy functional. Since we are interested in the liquid condensation from the vapor, we use non-conserved order parameter. Then the dynamics of phase transformation is governed by the interface-limited growth. Conserved order parameter would be more appropriated for the wetting by phase separation[17,18] where the dynamics of phase transformation is governed by the diffusion-limited growth.

The free energy $F[\psi]$ is written as the square-gradient form[10]

$$F[\psi] = \frac{1}{2}\int \left[ D(\nabla \psi)^2 + h(\psi) + V(r)\psi(r) \right] d\mathbf{r}. \qquad (2)$$

The local part $h(\psi)$ of the free energy functional $F[\psi]$ determines the bulk vapor-liquid phase diagram and the value of the order parameter $\psi$ in equilibrium phases. The wall potential $V(r)$ accounts for the long-range dispersion force from the wall of capillary. The local part of the free energy (grand potential) $h(\psi)$ we use[19,20] consists of two parts:

$$h(\psi) = h_0(\psi) + 6\Delta\mu\left(\frac{\psi^2}{2} - \frac{\psi^3}{3}\right) \qquad (3)$$

where $\Delta\mu$ is the chemical potential measured from the liquid-vapor coexistence (binodal). The bulk free energy $h_0(\psi)$ is given by

$$h_0(\psi) = C\psi^2(1-\psi)^2 \qquad (4)$$

where we assume that $\psi_v = 0$ represents the vapor phase, while $\psi_l = 1$ represents the liquid phase. $C$ is a constant which will be related to the bulk compressibility. The undersaturated vapor and the metastable liquid is characterized by a positive chemical potential $\Delta\mu > 0$, while the oversaturated vapor and the stable liquid is characterized by the negative chemical potential $\Delta\mu < 0$. Therefore, the free energy of the liquid phase $h(\psi_l = 1) = \Delta\mu$ is higher than the free energy of the vapor phase $h(\psi_v = 0) = 0$ when consider the capillary condensation and the vapor is undersaturated. Using the ideal gas expression, the chemical potential $\Delta\mu$ is approximately given by

$$\Delta\mu = k_B T \ln\left(\frac{p_{sat}}{p_{vap}}\right) \qquad (5)$$

where $p_{vap}$ is the vapor pressure and $p_{sat}$ is that of the saturated vapor. The capillary condensation is expected when $\Delta\mu > 0$ and the wetting transition occurs when $\Delta\mu \to 0^+$, while the coexistence of heterogeneous and homogeneous nucleation is expected when $\Delta\mu < 0$.[20] Then, the metastable liquid and the stable vapor phase in the undersaturated vapor pressure with $\Delta\mu > 0$ is represented by the free energy density (3) as a function of the order parameter $\psi$. Figure 1 shows the local part of the free energy $h(\psi)$ as the function of the order parameter $\psi$. The negative substrate potential pulls down the positive local minimum at $\psi_l = 1$ and the minimum becomes negative and is lower than the one at $\psi_v = 0$. Then the liquid phase is stable and liquid condensation occurs near the substrate.

Looking at eqs.(2) and (3), we can interpret the effect of the attractive substrate potential ($V(r)<0$) on the undersaturated vapor is to decrease the positive chemical potential $\Delta\mu>0$ toward the negative effective chemical potential according to

$$\Delta\mu \to \Delta\mu + V(r) \qquad (6)$$

Therefore, attractive substrate potential turns the undersaturated vapor into the effectively oversaturated vapor. Then, the substrate potential $V(r)$ promotes the nucleation (condensation) of liquid droplets near the substrate. Furthermore, this effectively positive chemical potential (6) becomes a driving force of liquid-vapor interface and the nucleus of liquid droplet can grow. This effectively positive chemical potential (6) thus determines the interfacial velocity and the morphology of liquid bridge.

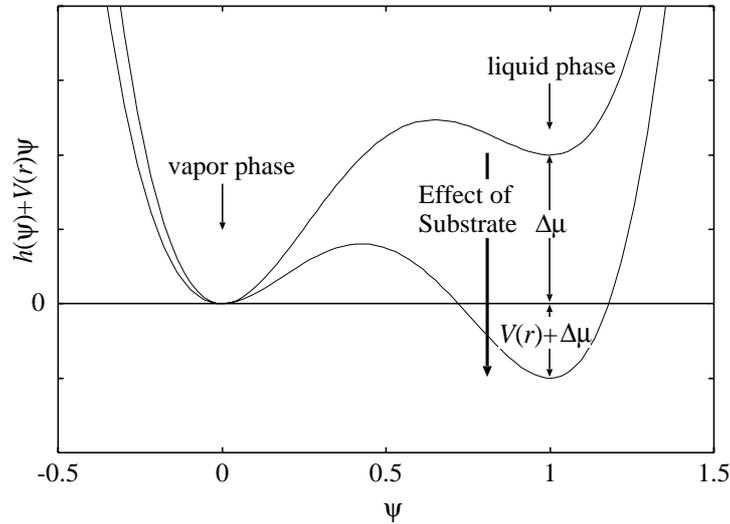

**Figure 1.** The local part of the free energy $h(\psi)$ as the function of the order parameter $\psi$. When the chemical potential is positive ($\Delta\mu>0$), the vapor is undersaturated and the local minimum at $\psi_v=0$ represents the stable vapor and another one at $\psi_l=1$ represents the metastable liquid. The metastable liquid phase becomes locally stable near the substrate due to the effect of substrate by the wall potential.

In this work we use the simplest intermolecular potential of the form $\sigma$

$$V(r) \propto -\left(\frac{\sigma}{r}\right)^6 \qquad (7)$$

where $r$ is the intermolecular distance and $\sigma$ is the effective size of molecular. Equation (7) gives the substrate potential for the homogeneous infinitely flat surface

$$V(z) = -\gamma_{ls}\left(\frac{1}{z/\sigma+1}\right)^3 \qquad (8)$$

where $z$ is the distance from the flat wall. The parameter $\sigma$ accounts for the range of potential. The liquids-substrate surface tension $\gamma_{ls}$ is related to the Hamaker constant.[1,5]

In order to promote the bridge formation, we use the chemically heterogeneous striped surface with width $2d$ shown in Figure 2. We consider a stripe that has finite width 2d along $x$-axis and infinite length along $y$-axis. Therefore we will consider effectively two-dimensional problem with cylindrical droplets or cylindrical bridges in this paper. Then the potential at $z$ from the substrate made by the lower substrate, for example, is given by[21,22]

$$V(x,z) = -\gamma_{ls}^{(1)}\left(\frac{1}{\tilde{z}+1}\right)^3 - \frac{\left(\gamma_{ls}^{(2)}-\gamma_{ls}^{(1)}\right)}{2}\left\{\frac{1}{(\tilde{x}-\tilde{d})^3} - \left(\frac{\tilde{r}_-}{(\tilde{x}-\tilde{d})(\tilde{z}-1)}\right)^3 + \frac{3}{2(\tilde{x}-\tilde{d})(\tilde{z}-1)\tilde{r}_-}\right\}$$

$$+ \frac{\left(\gamma_{ls}^{(2)}-\gamma_{ls}^{(1)}\right)}{2}\left\{\frac{1}{(\tilde{x}+\tilde{d})^3} - \left(\frac{\tilde{r}_+}{(\tilde{x}+\tilde{d})(\tilde{z}-1)}\right)^3 + \frac{3}{2(\tilde{x}+\tilde{d})(\tilde{z}-1)\tilde{r}_+}\right\} \qquad (9)$$

where all the lengths are scaled by $\sigma$ as $\tilde{z}=z/\sigma$, $\tilde{x}=x/\sigma$, $\tilde{d}=d/\sigma$, and $\tilde{r}_+ = \sqrt{(\tilde{x}+\tilde{d})^2+\tilde{z}^2}$, $\tilde{r}_- = \sqrt{(\tilde{x}-\tilde{d})^2+\tilde{z}^2}$. A similar artificial field called ghost field is used by Vishnyakov and Neimark[23] to initiate the liquid bridge formation. A similar striped surface is also used in microscopic simulations using molecular dynamics and Monte Carlo techniques.[24,25] Without using such an artificial field, a bridge could not be formed within a measurable time scale of the simulation. Then the adsorption occurs homogeneously on the whole substarate[11] and the evolution of the morphology of bridge cannot be traced.

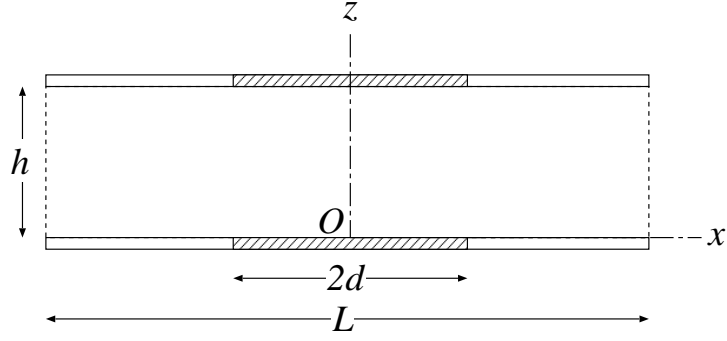

**Figure 2.** A model of striped surface and capillary. The stripe has width $2d$ and infinitely long along $y$ axis. Therefore we consider effectively two-dimensional problem in $x-z$ plane. Two stripes are facing each other within a parallel plate of the capillary. The width of slit is $h$ and the width of capillary is $L$ and the periodic boundary condition is imposed along $x$ axis.

**B. Thermodynamics of capillary condensation.** Before going into the results of the dynamics of bridge formation, we consider the thermodynamics of capillary condensation for the homogeneous wall. Total mean field free energy[26] for the liquid films of thickness $l$ with constant order parameter $\psi_l = 1$ adsorbed on single infinite substrates which consists of $x-y$ plane is given by

$$F_{wet} = \Delta\mu l + \int_0^l V(z)dz + \gamma_{lv} \qquad (10)$$

where $\gamma_{lv}$ is the liquid-vapor surface tension. It can be calculated from the interfacial profile obtained from the stationary solution $\delta F / \delta \psi = 0$ of the Ginzburg-Landau equation (1) at the two-phase coexistence ($\Delta\mu = 0$). In our model it is given by

$$\gamma_{lv} = \frac{1}{12}\sqrt{\frac{D}{2}} \qquad (11)$$

The second term of (10) of the integrated wall potential is also known as the disjoining potential.[1] By minimizing this free energy (10) with respect to the film thickness $l$, we obtain the film thickness $l_{min}$ determined from

$$\Delta\mu + V(l_{min}) = 0 \qquad (12)$$

where the first term is positive for the undersaturated vapor and the second term is negative as the wall is attractive. Equation (12) is the special form of the so-called augmented Young equation[27,28]

$$\gamma_{lv} \frac{l''(x)}{\sqrt{\left[1+(l'(x))^2\right]^3}} = \Delta\mu + V(l(x)) \qquad (13)$$

when the liquid vapor interface is flat $l(x) = l_{min}$. For the model wall potential (8), by solving eq. (12) we have

$$l_{min} = \sigma\left[\left(\frac{\gamma_{ls}}{\Delta\mu}\right)^{1/3} - 1\right] \qquad (14)$$

when $\gamma_{ls} > \Delta\mu$ otherwise $l_{min} = 0$ which means no wetting film on the substrate. The maximum undersaturation $\Delta\mu_{sp} = \gamma_{ls}$ defines the spinodal of the (first order) wetting transition. On the other hand when $\Delta\mu \to 0^+$, we will have a diverging wetting film thickness $l_{min} \to 1/\Delta\mu^{1/3} \to \infty$ which indicates the complete wetting.

In capillary wetting where the two substrates are separated by a distance $h$, we have to take into account the substrate potentials from two opposite substrates. Then the mean-field free energy is given by

$$F_{film} = 2\left[\Delta\mu l + \int_0^l \{V(z) + V(h-z)\}dz + \gamma_{lv}\right] . \qquad (15)$$

A factor 2 accounts for the two adsorbed films on the two substrates. Minimization of this free energy with respect to the film thickness $l$ gives the thickness of wetting film $l_{min}$ from

$$\Delta\mu + V(l_{min}) + V(h-l_{min}) = 0 \qquad (16)$$

similar to (12) and gives the free energy of two separated wetting films $F_{film}$, which should be compared with the free energy of the capillary-condensed phase

$$F_{film} = 2\left[\Delta\mu\left(\frac{h}{2}\right) + \int_0^{h/2}\{V(z) + V(h-z)\}dz\right] \qquad (17)$$

with the film thickness $l = h/2$. The capillary condensation occurs if $F_{cc} < F_{film}$. A rough estimation of bridge formation occurs

$$l_{\min} = \sigma\left[\left(\frac{\gamma_{ls}}{\Delta\mu}\right)^{1/3} - 1\right] \cong \frac{h}{2} \qquad (18)$$

when the two droplets touch each other. We observe from eqs. (12) and (16) that the growth of wetting film closely follows the equi-potential contour of substrate potential $V(z)$, which will be confirmed in the next section.

### III. Numerical Results

**A. Cell dynamics method.** In this section, we show the numerical results of simulation. We do not solve the time-dependent Ginzburg-Landau (TDGL) equation directly. Instead, we use the cell dynamics method[29] to solve TDGL approximately. In a previous study we[20,30] have already shown that the cell dynamics method is accurate enough and numerically very efficient to study the general feature of the dynamics of various first order phase transformations.

In the cell dynamics method, both time and space are discretized. The differential evolution equation (1) is transformed into a finite difference equation of the form

$$\psi(t+1,n) = F[\psi(t,n)] + \xi(t,n) \qquad (19)$$

where the time $t$ is discrete and an integer, and the space is also discrete and is expressed by the integral site index $n$. The general form of the mapping function $F[\psi]$ can be found in the reference[20]. The traditional tanh mapping[29] is not used. Rather we use the mapping directly derived from the free energy functional (2)-(4),[30,31] which is essential for studying the subtle nature of nucleation and growth when one phase is metastable and the other is stable. The thermal noise $\xi(t,n)$ is added to (1) in (19) which is related to the absolute temperature $T$ from the fluctuation-dissipation theorem

$$\langle \xi(t,n)\xi(t',n')\rangle = k_B T \delta_{t,t'}\delta_{n,n'}. \qquad (20)$$

In this paper, we will use a uniform random number ranging from $-\xi_0$ to $+\xi_0$.[20,29] Then, the parameter $\xi_0^2$ represents the absolute temperature since

$$\xi_0^2 \propto T. \qquad (21)$$

The temperature $T$ plays a role only through the thermal noise $\xi_0^2$.

Further detail of the numerical method to solve the finite difference equation and its boundary condition can be found in our previous study of heterogeneous nucleation.[20] The differences between this study and the previous one are the introduction of a long-range dispersion force (8) and (9) and the sign of chemical potential $\Delta\mu$. The vapor phase confined within the capillary is stable in capillary condensation, while it is metastable and should transform into liquid in heterogeneous nucleation. The crucial role played by this long-range potential will be clarified later in this section. In order to reduce the numerical work further, in this article, we consider only a two-dimensional system. Therefore we consider a section of infinite striped surface.

**B. Growth of a single droplet.** In order to study the evolution of a single droplet, we first consider the cases when only the lower substrate is striped. The upper substrate consists of a homogeneous neutral wall with $\gamma_{ls}^{(1)} = \gamma_{ls}^{(2)} = 0$.

Figure 3 shows the evolution of droplet on a striped single substrate calculated using the cell dynamics method. We consider two cases; (i) the lower substrate consists of neutral substrate with $\gamma_{ls}^{(1)} = 0$ and attractive stripe of the width $2d = 201$ with $\gamma_{ls}^{(2)} = 0.25$, and (ii) the lower substrate consists of weakly attractive substrate with $\gamma_{ls}^{(1)} = 0.02$ and attractive stripe of the width $2d = 201$ with $\gamma_{ls}^{(2)} = 0.25$. Therefore the contact angle $\theta$ defined by Young's equation

$$\gamma_{lv}\cos\theta = \gamma_{vs} - \gamma_{ls} \tag{22}$$

is $\theta = \pi/2$ on the neutral substrate and $\theta = \cos^{-1}(0.02/0.042) \cong 62°$ on the weakly attractive substrate as $\gamma_{vs} = 0$ and $\gamma_{lv} = 1/24 \cong 0.042$. The contact angle $\theta$ of the stripe is $\theta = 0°$ since $\cos\theta = 0.25/0.042 >> 1$.

The potential parameters $C$ in (4) and $D$ in (2) are fixed to $C = 1/4$ and $D = 1/2$ respectively throughout this article. The undersaturation $\Delta\mu$ and the absolute temperature $T \propto \xi_0^2$ are fixed to $\Delta\mu = 0.0125$ and $T \propto \xi_0^2 = 1/80$ for both case (i) and (ii). The simulation box is fixed to $513 \times 128$, which is large enough to capture the essential feature of the evolution. Therefore the width of capillary is $h = 128$.

The range of potential $\sigma$ is fixed to $\sigma = 50$ so that the size of droplet from (14) becomes comparable to the simulation box.

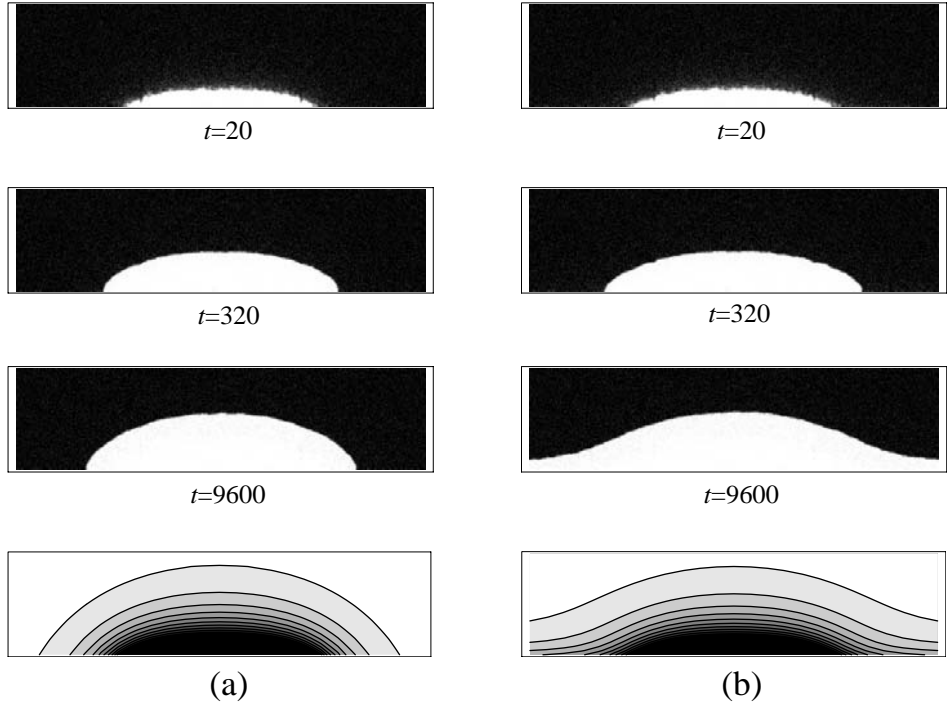

**Figure 3.** Evolution of the droplet on nano-striped substrates and the contour plot of the wall potentials of the stripe and the substrate when (a) $\gamma_{ls}^{(1)} = \gamma_{ls}^{(2)} = 0$ at the top and, $\gamma_{ls}^{(1)} = 0$ and $\gamma_{ls}^{(2)} = 0.25$ at the bottom (case (i)), and (b) when $\gamma_{ls}^{(1)} = \gamma_{ls}^{(2)} = 0$ at the top and $\gamma_{ls}^{(1)} = 0.02$ and $\gamma_{ls}^{(2)} = 0.25$ at the bottom (case (ii)). The morphology of the droplet is different from the circular shape predicted from the macroscopic capillary theory as we cannot neglect wall potential. In case (ii) in Figure 3(b), the droplet grows not only vertically but also laterally. Contour plots of the wall potential at the bottom created by the lyophilic stripe and the substrates are similar to the growing front of the droplet.

The initial morphology of droplets for two cases (i) and (ii) in Figure 3 seems quite similar to those observed experimentally by atomic force microscopy.[28] It differs significantly from the semi-circular shape predicted from the macroscopic capillary theory that neglects the wall potential $V(z)$ called disjoining potential. The morphology is directly related to the contour of wall potential as we can expect from eq. (12) for the homogeneous wall. For the droplet on the striped surface, the equilibrium shape of the wetting layer would be obtained by an equation similar to (12) by minimizing the total free energy. Unfortunately, we cannot obtain the analytical expression for the total free energy and, therefore, the equilibrium equation similar to (12). But, we can expect that the equation that corresponds to (12) would be similar to the well-know augmented Young equation (13). The curvature dependence of the liquid-vapor surface tension which can be obtained directly from the minimization of total free energy (2) in principle will implicitly include the capillary force (the first term of (13)). Therefore, our cell-dynamics simulation can simulate the morphology of nano-droplet on the substrate directly without resorting to the semi-phenomenological augmented Young equation (13).

There is also the effect of the attractive potential of the stripe even outside the stripe as the long-range dispersion force is originally isotropic. Then, the vapor near the substrate is affected by this attractive potential. The liquid starts to condense near the substrate even outside the boundary of the stripe. Then the droplet edges are not pinned at the boundary of the stripe. Rather the droplet spread out from the stripe and pinned somewhere on the neutral substrate (Fig. 3(a)). Such a spillage of the droplet from the boundary of the stripe has already been imagined as a character of nano-fluids.[27]

When there is a weak attractive wall potential even outside the stripe (case (ii)), it is expected that the droplet will spread over the whole surface of the substrate. Figure 3 (b) clearly shows that the evolution and spreading of the droplet occurs simultaneously when $\gamma_{ls}^{(1)} = 0.02$ and $\gamma_{ls}^{(2)} = 0.25$. Initially the droplet starts to nucleate on the stripe. Later, the droplet grows not only vertically but also laterally. In contrast to case (a), the base of the droplet does not stop but seems to grow infinitely.

The contours plot of the wall potential made by the stripe and the substrate is also shown in Figure 3. The growing front of the droplet closely follows the equi-potential line. This can be easily explained from (12) if the capillary force due to the curvature of the meniscus could be neglected. The weak curvature dependence of the liquid-vapor surface tension will bring the rounding of the moving front. Then, the velocity of curved meniscus is approximately the velocity of planer front. The steady state front velocity $v$ for the TDGL is approximately given by[20,32]

$$v \approx -3\sqrt{\frac{D}{2C}}(\Delta\mu + V(z)) \qquad (16)$$

where $z$ is the distance from the wall. Since the vapor is undersaturated, the chemical potential difference is positive $\Delta\mu > 0$. However, the attractive (negative) wall potential $V(z) < 0$ makes the effective chemical potential $\Delta\mu + V(z)$ negative. Then the liquid droplet grows as far as $\Delta\mu + V(z) < 0$ near the wall since the growth velocity $v$ is positive ($v > 0$). The growth stops when $\Delta\mu + V(z) \approx 0$ which corresponds to the equilibrium thickness of the wetting layer determined from (12). Therefore the morphology of the droplet or the height of the droplet $z$ is determined from the equipotential contour line with $V(z) \approx \Delta\mu$. Then the contour plot of the wall potential roughly traces the morphology of the droplet (Figure 3).

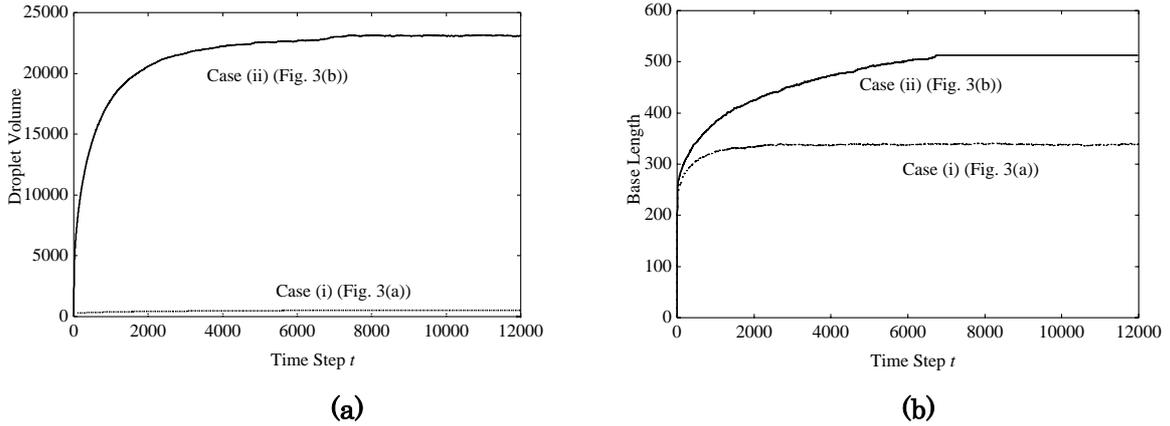

**Figure 4.** Evolution of (a) droplet volume and (b) base length as the functions of time step for the case (i) and case (ii) of Figure 3. The base length is longer than the stripe width $2d = 200$.

Figure 4 shows the evolutions of droplet volume (Figure 4(a)) and base length (Figure 4(b)) for the case (i) and (ii). In case (i), saturation of volume (Figure 4(a)) and base length (Figure 4(b)) is observed. However, the base length is larger than the width of the stripe ($2d = 201$). Therefore, the droplet edge is not pinned by the boundary of the stripe. Rather the droplet liquid spills out from the stripe,[27] and makes the contact angle $\theta \cong \pi/2$ as the liquid-substrate surface tension is $\gamma_{ls}^{(1)} = 0$. The morphology and spillage of the droplet from the stripe is due to the long ranged attractive potential from the stripe, which could be inferred from the shape of the contour of the wall potential in Figure 3 (a). In case (ii) the volume and the base length increase indefinitely due to the weakly attractive substrate potential which causes the growth as well as the spreading of the droplet (Figure 4(b)) with contact angle $\theta \approx 62°$. The saturation of volume and base length after $t \geq 7000$ is due to the finite size of simulation box ($513 \times 128$). Since we are studying the droplet at a finite temperature $T \propto \xi_0^2 = 1/80$, the droplet surface fluctuates due to the capillary fluctuation. Thus the droplet volume shows a small fluctuation in Figure 4.

In order to study the temperature dependence, we also calculated the evolutions of droplet volume and base length for higher temperatures. As expected, the temperature only plays a role at the initial stage of nucleation. The evolution curve of droplet volume and base length for higher temperatures $T \propto \xi_0^2 = 1/50$ and $\xi_0^2 = 1/10$, for example, cannot be distinguished from Figure 3 for $T \propto \xi_0^2 = 1/80$. The effect of temperature is considered to be marginal. It only provides the seeds to start the nucleation at the surface of the stripe. Therefore, in contrast to the usual phase transformation in a solid where both the steady state nucleation and the growth of nucleated grain determine the whole dynamics of phase transformation,[33] only the interface-limited growth of the droplet meniscus determines the evolution of the droplet. Then, the time scale of phase transformation should be relatively independent of the temperature since the interface velocity $v$ in eq. (16) does not depend on the temperature at all.

A similar but totally different phenomenon of critical droplet formation on a substrate due to the heterogeneous nucleation has been studied by several authors.[20, 34-36] Since we are interested in the wetting phenomena where the liquid is volatile and the vapor is undersaturated with positive chemical potential $\Delta\mu > 0$, the droplet we considered in this subsection is stable. On the other hand, the critical droplet of heterogeneous nucleation is in the transient and metastable state. Therefore, it will grow without bound as the vapor is already oversaturated with negative chemical potential $\Delta\mu < 0$. The substrate only plays a secondary role that initiates nucleation of droplets in heterogeneous nucleation,[20] while it plays a crucial role that supports stable droplet of wetting layer in wetting phenomena.

**C. Growth of liquid bridge.** Next we consider the cases when the upper substrate as well as the lower substrate is striped (Figure 2). Figure 5 shows the evolution from drops to bridge when the two stripes are facing each other within the parallel plate of the capillary. The upper and the lower substrate consist of neutral substrate with $\gamma_{ls}^{(1)} = 0$ and attractive stripe of the width $2d = 201$ with $\gamma_{ls}^{(2)} = 0.25$. These values for the surface tension $\gamma_{ls}^{(1)}$ and $\gamma_{ls}^{(2)}$ are exactly the same as those used to study the droplet growth in Figure 4. We considered two cases when (i) weak undersaturation $\Delta\mu = 0.0125$ and (ii) strong undersaturation $\Delta\mu = 0.025$. The absolute temperature $T \propto \xi_0^2$ are fixed to $T \propto \xi_0^2 = 1/80$ for both (i) and (ii). The simulation box is $513 \times 218$, and the range of potential $\sigma$ is $\sigma = 50$ again. Therefore the width of capillary is $h = 128$.

Initially the growth of the droplet is similar to Figure 4. However the growing two droplets merge at midway between two substrates and continue to grow. Eventually they form stable bridge with almost constant size. The size of bridge finally formed depends on the undersaturation $\Delta\mu$. The size becomes larger as we decrease $\Delta\mu$, which is expected from eq. (12) and the contour plots of the wall potential created by the stripe at the bottom of Figure 4(a).

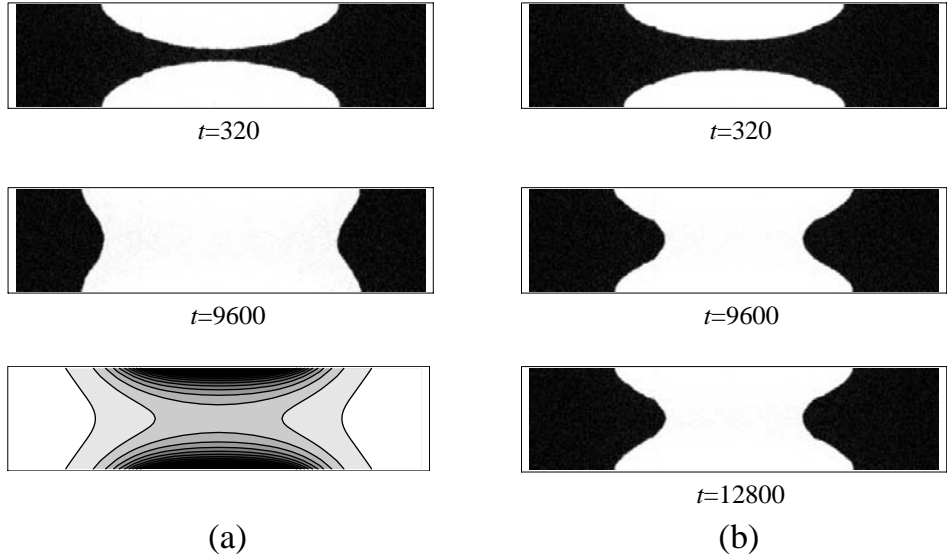

(a)            (b)

**Figure 5.** Transformation of droplets to a bridge within a nano-striped nano-capillary (a) for the case (i) $\Delta\mu = 0.0125$, and (b) for the case (ii) $\Delta\mu = 0.025$. The morphologies of the droplet and the bridge are different from circular segment predicted from the macroscopic capillary theory. Contour plots of the wall potential created by the stripe at the bottom of (a) closely follow the growing fronts and final morphology of the bridge.

The morphologies of the droplets and the bridges certainly deviate from the circular shape. The liquid-vapor interface does not consist of the segment of circle with constant curvature. Initially, the curvature is not concave but is convex when two droplet start to merge. The classical capillary theory predicts a concave curvature because pressure of stable vapor is higher than the metastable liquid from the Young-Laplace equation. Such a concave curvature is observed only at the later stage of evolution ($t = 9600$ and $t = 12800$ in Figure 5). The final stable bridges mainly consist of expected concave curvature meniscus. However, the meniscus deviates significantly from the circular shape near the substrate due to the substrate potential. Similar to the case of droplet, the volume and the base length of the bridge seems to saturate as the bridge edge is pinned at somewhere outside the stripe. Again, the base length is larger than the width of stripe ($2d = 201$).

Figure 6 shows the evolution of the bridge volume (Figure 6(a)) and base length (Figure 6(b)). The bridge volume for the lower undersaturation ($\Delta\mu = 0.0125$) is much larger than that for the higher undersaturation ($\Delta\mu = 0.025$). The base length is also slightly larger for the former than the latter.

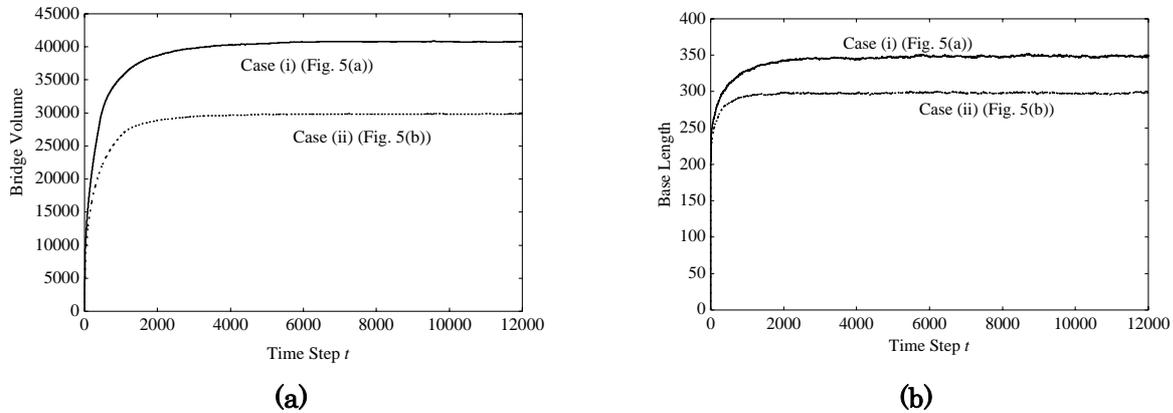

**Figure 6.** Evolution of (a) the bridge volume, and (b) its base length as the functions of time step for the case (i) $\Delta\mu = 0.0125$ and case (ii) $\Delta\mu = 0.025$. The base length is longer than the stripe width $2d = 201$. The size and base length becomes larger for smaller undersaturation $\Delta\mu$.

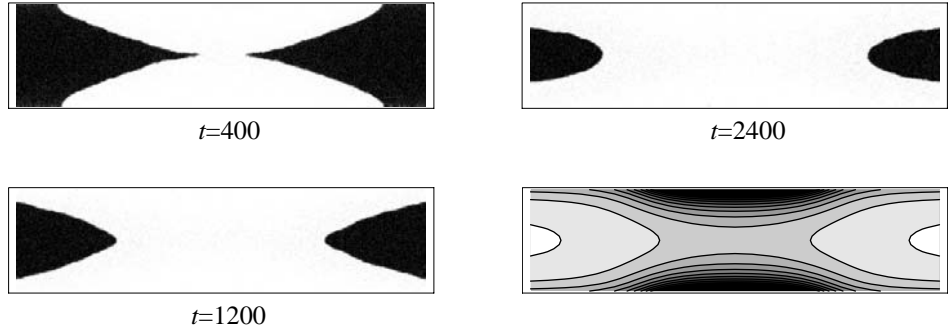

**Figure 7.** The same as Figure 5 when the substrate is also weakly wet with $\gamma_{ls}^{(1)} = 0.02$ and $\gamma_{ls}^{(2)} = 0.25$. The other parameters are the same as those used in Figure 3 (b). The liquid bridge continues to grow to form a vapor bubble.

Finally we consider the case when the whole substrate is attractive. Figure 7 shows the evolution of bridge when $\gamma_{ls}^{(1)} = 0.02$ and $\gamma_{ls}^{(2)} = 0.25$. Those parameters and other parameters $\Delta\mu$ and $\xi_0^2$ are the same as in Figure 3(b). We found that the two droplets grow not only vertically but also laterally. After these two droplets merge, the whole bridge still continues to grow to form a bubble of vapor phase. This vapor bubble continues to shrink and whole space of capillary is finally occupied by the liquid phase. This is exactly the scenario of capillary condensation proposed by Everett and Heynes.[14] Our cell dynamics simulation can successfully confirm the dynamics of capillary bridging. It must be noted, however, that the droplet formation is assumed to occur spontaneously from the fluctuation of wetting film[14]. In our simulation such a fluctuation is artificially induced by the striped wall potential to extract the evolution of a single bridge. Otherwise, the evolution of multiple droplets and bridges overlap[20], and we cannot separate the evolution of single bridge.

As in the case of droplet formation in the previous subsection, the temperature plays role only at the initial stage of droplet formation, and whole process of bridge formation does not depend on the temperature as the dynamics is governed by the interface-limited growth with temperature-independent interfacial velocity (16). Therefore, the time scale of bridge formation in our model is relatively independent of the temperature. This result contradicts a similar theoretical calculation by Restagno et al.[10] and an experimental result by Szoszkiewicz and Riedo.[8] Both authors observed thermally activated bridge nucleation with the nucleation time $\tau$ of the activation form $\tau \propto \exp(\Delta\Omega/k_BT)$, where $\Delta\Omega$ is the activation (nucleation) barrier. In our model, the nucleation of bridge is certainly thermally activated, too. However, the nucleation time is so short that the whole time scale of bridge formation is determined by the almost temperature-independent growth time. One reason of this discrepancy could be due to the use of our artificial striped geometry. Without such an artificial field, the incubation time[33] before the start of nucleation would be very long which, in practice, could determine the total bridge formation time. This story will explain the activation-type temperature dependence of bridge-formation time[8,10] that, in fact, is the incubation time.

We have also conducted a number of simulations by changing stripe width, capillary widths, and the strength of wall potential, etc. However, the qualitative results can be easily interpreted by equation (12).

## IV. Conclusion

In this article we have considered the dynamics and morphology of the nanoscopic liquid droplet and bridge within the TDGL model. It has been demonstrated that the dynamics of drop and bridge formation is driven by the long range attractive substrate potential. In particular the front velocity of transformed liquid phase is determined from the balance of the positive chemical potential of the undersaturated vapor and the negative wall potential. The morphology is determined when the liquid-vapor front velocity vanishes, and the liquid-vapor interface stops. This corresponds to the augmented Young equation. Therefore the morphology of the liquid droplet and bridge within a nanoscopic capillary is determined not only from the macroscopic capillary force due to liquid-vapor interfacial tension but also from the contour of the substrate potential.

In a macroscopic droplet and bridge, the macroscopic capillary force that is represented by the left hand side of (13) dominates and the substrate potential $V(l(x))$ is negligible. Then the macroscopic capillary theory can describe the morphology of droplets and bridges. In a nanoscopic droplet and bridge, our cell dynamics simulation clearly demonstrates that the substrate potential frequently called the disjoining potential is not negligible. Since the long range molecular force (8) plays role only at nano-scale, the morphology and dynamics of such nanoscopic fluid sensitively depend on those molecular forces in nanometer range rather than on the macroscopic capillary force. A similar conclusion was derived for the motion of nanodroplets near edges and wedges of a geometrically heterogeneous substrate.[37]

Although, we have studied the two-dimensional system with cylindrical droplets or bridges, our cell dynamics method can easily be extended to three-dimensional systems. However, results for three-dimensional systems can be easily inferred from our result for two-dimensional system. For example, for the circular stripes, we will have axially symmetric droplets or bridges whose morphology will be determined from the equi-potential contour of wall potential again. For the linear stripe, the results will be similar to our results if the length of stripe is sufficiently long. Again, the morphology is mainly determined from the wall potential for nano-scale droplets or bridges.

Since the TDGL model with the cell dynamics method is flexible and computationally efficient, yet it has a direct connection to the thermodynamics of the system considered, it can be used to study various scenarios of phase transformation.[20,30] Although the lattice gas model[11-13] with Monte Carlo method is frequently used to study the phase transformation of nano-fluids, they can be mainly used to study the equilibrium properties. On the other hand, in our cell dynamics method, the dynamics is driven by the free energy and the thermal noise. Therefore, our TDGL model with cell dynamics method is more natural and flexible choice to study various scenarios of condensation phenomena, in particular, the time evolution including capillary condensation and evaporation.

**References**

(1) Israelachvili, J. N. *Intermolecular and surface forces*, 2nd ed.; Academic Press; London, **1992**.

(2) Evans, R. *J. Phys.: Condens. Matter* **1990**, 2, 8989.

(3) Derjaguin, B. V.; Churaev, N. V. *J. Colloid Interface Sci.* **1976**, 54, 157.

(4) Iwamatsu, M.; Horii, K. *J. Colloid Interface Sci.* **1996**, 182, 400.

(5)  Butt, H-J.; Graf, K.; Kappl, M. *Physics and Chemistry of Interface*; Wiley-VCH: Weiheim, **2003**.

(6) Bocquet, L.; Charlaix, E.; Ciliberto, S.; Crassous, J. *Nature* **1998**, 396, 735.

(7) Kohonen, M. M.; Maeda, N.; Christenson, H. K. *Phys. Rev. Lett.* **1999**, 82, 4667.

(8) Szoszkiewicz, R.; Riedo, E. *Phys. Rev. Lett.* **2005**, 95, 135502.

(9) Schmidt, I.; Binder, K. *Z. Phys. B*, **1987**, 67, 369.

(10) Restagno, F.; Bocquet, L.; Biben, T. *Phys. Rev. Lett.* **2000**, 84, 2433.

(11) Lum, K.; Luzar, A. *Phys. Rev. E* **1997**, 56, R6283.

(12) Leung, K.; Luzar, A.; Bratko, D. *Phys. Rev. Lett.* **2003**, 90, 065502.

(13) Jang, J.; Ratner, M. A.; Schatz, G. C. *J. Phys. Chem. B* **2006**, 110, 659.

(14) Everett, D. H.; Haynes, J. M. *J. Colloid Interface Sci.* **1972**, 38, 125.

(15) Talanquer, V.; Oxtoby, D. W. *J. Chem. Phys.* **2001**, 114, 2793.

(16) Valencia, A.; Brinkmann, M.; Lipowsky, R. *Langmuir* **2001**, 17, 3390.

(17) Law, B. M. *Prog. Surf. Sci.* **2001**, 66, 159.

(18) Bonn, D.; Bertrand, E.; Meunier, J.; Blossy, R. *Phys. Rev. Lett.* **2000**, 84, 4661.

(19) Jou, H-J.; Lusk, M. T. *Phys. Rev. B* **1997**, 55, 8114.

(20) Iwamatsu, M. *J. Chem. Phys.* **2007**, 126, 134703.

(21) Cole, M. W.; Vittoratos, E. *J. Low. Temp. Phys.* **1976**, 22, 223.

(22) Bauer, C.; Dietrich, S. *Phys. Rev. E* **1999**, 60, 6919.

(23) Vishnyakov, A.; Neimark, A. V. *J. Chem. Phys.* **2003**, 119, 9755.

(24) Gac, W.; Patrykiejew, A.; Sokolowski, S. *Surf. Sci.* **1994**, 306, 434.

(25) Vishnyakov, A.; Piotrovskaya, E. M.; Brodskaya, E. N. *Adsorption* **1998**, 4, 207.

(26) Gartica, S. M.; Calbi, M. M.; Cole, M. W.; *Phys. Rev. E* **1999**, 59, 4484.

(27) Dietrich, S.; Popescu, M. N.; Rauscher, M. *J. Phys. Condens. Matter* **2005**, 17, S577.

(28) Checco, A.; Gang, O.; Ocko, B. M. *Phys. Rev. Lett.* **2006**, 96, 056104.

(29) Oono, Y.; Puri, S. *Phys. Rev. A* **1988**, 38, 434.

(30) Iwamatsu, M.; Nakamura, M. *Jpn. J. Appl. Phys. Pt 1.* **2005**, 44, 6688.

(31) Ren, S.R.; Hamley, I.W. *Macromolecules* **2001**, 34, 116

(32) Chan, S-K. *J. Chem. Phys.* **1977**, 67, 5755.

(33) Christian, J. W. *The theory of transformations in metals and alloys: an advanced textbook in physical metallurgy*; Pergamon Press; Oxford, **1965**, Chapter XI.

(34) Talanquer, V.; Oxtoby, D. W. *J. Chem. Phys.* **1996**, 104, 1483.

(35) Valencia, A.; Lipowsky, R. *Langmuir* **2004**, 20, 1986.

(36) Granasy, L.; Pusztai, T.; Saylor, D.; Warren, J. A. *Phys. Rev. Lett.* **2007**, 98, 035703.

(37) Moosavi, A.; Rauscher, M.; Dietrich, S. *Phys. Rev. Lett*. **2006**, 97, 236101.